\documentstyle[prb,aps,twocolumn,epsf,floats]{revtex}

\begin{document}

\wideabs{
\title{Simple technique for superconducting joints quality estimation in bulk melt-processed high temperature superconductors}
\author{A. A. Kordyuk, V. V. Nemoshkalenko, A. I. Plyushchay}
\address{Institute of Metal Physics, 36 Vernadsky St., Kyiv 03142, Ukraine}
\author{T. A. Prikhna}
\address{Institute for Superhard Materials, 2 Avtozavodskaya St., Kyiv 04074, Ukraine}
\author{W. Gawalek}
\address{Institut f\"ur Physikalische Hochtechnologie, Winzerlaer Str. 10, D-07743, Jena, Germany}
\date{\today}
\maketitle

\begin{abstract}
We propose an empirical approach to estimate the quality of superconducting joints (welds) between blocks of bulk high temperature superconductors (HTS). As a measuring value, we introduce a joint's quality factor and show its natural correlation with joint's critical current density. Being simple and non-destructive, this approach is considered to be quite important to solve the problem of utilization of HTS in large scale applications. The approach has been applied to characterize the joint's quality of melt-processed Y-123 joined by Tm-123 solder. 
\end{abstract}
\pacs{7460Jg, 7480Bj, 8525Ly}
}

Bulk melt-processed high temperature superconductors (MP HTS) have been appeared to be quite suitable for such large scale HTS applications as superconducting motors, contactless bearings, flywheels for energy storage, levitation transport.\cite{Hull,Campbell} Remarkable values of critical current densities have been achieved for these quasi-single crystals and the main problem now, which is an obstacle in the way of their practical utilization, is superconducting joining of them into appropriate for this utilization blocks.\cite{IRC,Arg1} While the problem being under active investigation \cite{IRC,Arg1,Arg2,Prikhna,Arg3} has not been completely resolved yet, it is vitally important to find a simple method for quantitative characterization of superconducting joints. 

The most physical way to describe a superconducting joint quantitatively is to determine density of critical current that flows through it. While the transport measurements are quite difficult to exploit here, the contactless techniques appear to be the most attractive.\cite{Arg2,Arg3} It has been shown that the levitation force measurements,\cite{Arg1} magneto-optical imaging techniques,\cite{Arg1} or Hall-probe magnetometry\cite{Arg2,Prikhna} can be successfully used to characterize superconducting joint's quality. At last, in a very recent paper,\cite{Arg3} an approach to estimate critical current density through superconducting joint for ring samples was proposed.

In this Letter we propose a simple non-destructive contactless method to evaluate the quality of superconducting joint and estimate critical current density through it. The method is based on the technique of critical current density determination from local levitation force measurements which, as has been shown in Ref. \onlinecite{APL2}, allows to determine critical current density in a thin undersurface layer. 

\begin{figure}[!b]
\begin{center}
\epsfxsize=8.47cm
\epsfbox{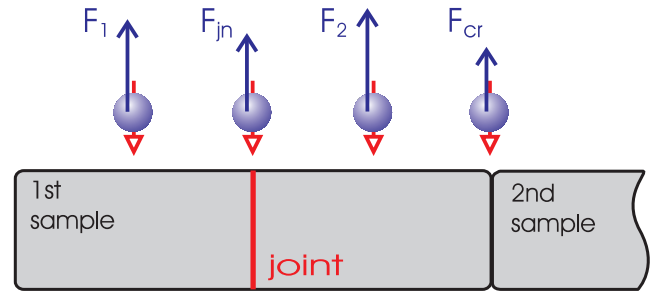}
\end{center}
\caption{A schematic illustration of 'four points procedure'. The levitation force, acting from screening current in a small spherical permanent magnet with vertical direction of magnetic moment, is measured in four points: above the 'uniform' parts of sample, $F_{1}$ and $F_{2}$ ($F_{un} = (F_{1}+F_{2})/2$), above the superconducting joint, $F_{jn}$, and above the 'crack' (mechanical joint of two samples), $F_{cr}$.}
\label{forces}
\end{figure}

We test the method on Y-123 MP HTS samples with superconducting joints made by solidification technique ("soldered joints") with Tm-123 as a solder. The details of the technique are described in Ref. \onlinecite{Prikhna}. Two types of experimental procedure were used: 'three times procedure' and 'four points procedure'. Within 'three times procedure' we measure the levitation force that acts on a small permanent magnet (PM) three times at one point: above the uniform sample before cutting, $F_{un}$, above the 'crack' (mechanical joint) after cutting, $F_{cr}$, and above the superconducting joint after soldering procedure, $F_{jn}$. 'Four points procedure' (see Fig.\ref{forces}), being less precise due to samples' inhomogeneity, was used for the samples which were not measured before cutting and joining, and $F_{un}$ was estimated as $(F_{1}+F_{2})/2$. We used two PMs: spherical one, of 1.5 mm in diameter with magnetic moment $\mu$ = 1.9 G cm$^3$, and cylindrical one, of 6.3 mm in diameter and 2.3 mm in width with $\mu$ = 38 G cm$^3$. The readout distances between centers of the magnets and HTS surface were 1.25 mm and 6 mm respectively. The basic experimental setup and the technique for critical current density determination have been described in details in Ref. \onlinecite{APL2}.

Based on such local levitation force measurements, one can introduce a joint's quality factor  
\begin{equation}
q = \frac{F_{jn} - F_{cr}}{F_{un} - F_{cr}}.
\label{qfac}
\end{equation}

This formula was constructed to satisfy the natural asymptotic conditions: $q \to 0$ when $F_{jn} \to F_{cr}$ (non-superconducting joint), and $q \to 1$ when $F_{jn} \to F_{un}$ (an ideal superconducting joint). In the following we will find the physical meaning of thus introduced quality factor through its relation with joint's critical current density $J_{jn}$.

\begin{figure}[!t]
\begin{center}
\epsfxsize=5cm
\epsfbox{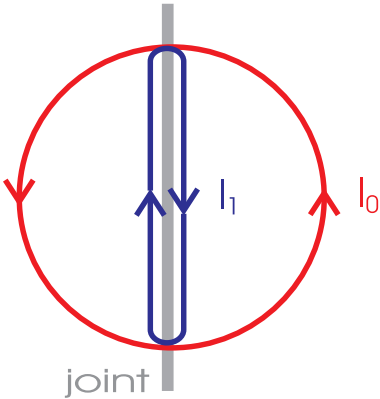}
\end{center}
\caption{A schematic illustration on currents circulating in HTS bulk which produce the levitation force.}
\label{currents}
\end{figure}

Let us consider the next naive picture of circular currents which are sketched at Fig.\ref{currents}. In a uniform sample a circular current $I_0$ flows mainly over the top surface of the sample \cite{APL2} and one can write $F_{un} = \alpha I_0$. If we crack the sample, the current (having the same value, until the size of the crack is much less than sample dimensions) cannot flow though the crack and has to close over the crack walls but one can consider this as an additional opposite ($I_1 = -I_0$) circular current flowing through the crack and over the crack walls: $F_{cr} = \alpha I_0 + \beta I_1 = (\alpha - \beta) I_0$. Next, it is quite natural to assume that for a sample with a superconducting joint one can estimate the real current through the joint as 
\begin{equation}
I_{jn} = I_0 - I_1 = \frac{J_{jn}}{J_c}I_0,
\label{Ijn}
\end{equation}
where $J_c$ is the bulk critical current density. Then $I_1 = (1 - (J_{jn}/J_c)) I_0$ and $F_{jn} = (\alpha - \beta (1 -  (J_{jn}/J_c))) I_0$. Substituting this into (1) we obtain 
\begin{equation}
q = \frac{J_{jn}}{J_c}.
\label{qJ}
\end{equation}

Thus, the superconducting joint's quality factor introduced by (\ref{qfac}) represents the ratio of density of the critical current that flows through the joint to the bulk critical current density of the sample. The Eq.\ref{qfac} by itself shows a simple way to estimate this quality factor but in combination with the method of $J_c$ determination from levitation force measurements \cite{APL2} gives a simple non-destructive method for determination of density of the critical current through the superconducting joint.

\begin{table}[!b]
\squeezetable
\caption{Critical current densities and quality factors of soldered joints of MP YBCO samples.}
\begin{tabular}{cccccccc}
Sample & $T_{tr}$, K& top & $J_{un}$, & $J_{jn}$, & $q_I$ & $q_{II}$ & $q_M$ \\
no.&    & plane & A/cm$^{2}$ & A/cm$^{2}$ & & \\ [1pt]
\hline \\
1 &1257 &  $\parallel$ ab & 1.5$\cdot$10$^4$ &1.2$\cdot$10$^4$ & 0.79 & 0.65 & 0.40 \\
   &          &  $\perp$ ab      & 7.6$\cdot$10$^3$ & 3.9$\cdot$10$^3$ & 0.52 & --       & 0.29 \\ [3pt]
2 &1257 &  $\parallel$ ab & 1.3$\cdot$10$^4$ & 3.2$\cdot$10$^3$ & 0.25 & 0.32 & 0.56 \\
   &          &  $\perp$ ab      & 7.2$\cdot$10$^3$ & 5.7$\cdot$10$^3$ & 0.79 & --       & 0.67 \\ [3pt]
3 &1253 &  $\parallel$ ab & 1.3$\cdot$10$^4$ & 1.6$\cdot$10$^3$ & 0.13 & 0.28 & 0.47 \\
   &          &  $\perp$ ab      & 9.1$\cdot$10$^3$ & 7.7$\cdot$10$^3$ & 0.84 & --       & 0.46 \\ [3pt]
4 &1253 &  $\parallel$ ab & 1.6$\cdot$10$^4$ & 8.5$\cdot$10$^3$ & 0.53 & 0.49 & 0.33 \\
   &          &  $\perp$ ab      & 9.3$\cdot$10$^3$ & 6.7$\cdot$10$^3$ & 0.72 & --       & 0.30 \\ [3pt]
5 &1263 &  $\parallel$ ab & 1.5$\cdot$10$^4$ & 7.2$\cdot$10$^3$ & 0.47 & 0.55  & 0.77 \\
   &          &  $\perp$ ab      & 8.2$\cdot$10$^3$ & 3.9$\cdot$10$^3$ & 0.47 & --       & 0.53 \\ [3pt]
6 &1263 &  $\parallel$ ab & 2.2$\cdot$10$^4$ & 9.4$\cdot$10$^3$ & 0.43 & 0.46 & 0.45 \\
   &          &  $\perp$ ab      & 7.1$\cdot$10$^3$ & 6.1$\cdot$10$^3$ & 0.85 & --       & 0.57 \\
\end{tabular}
\label{table}
\end{table}

Though the given derivation of Eq.\ref{qJ} from Eq.\ref{qfac} is quite simple, it has some restrictions which should be considered.  The first of them is a validity of consideration of levitation forces to be proportional to corresponding currents - we have assumed above that the introduced coefficients $\alpha$ and $\beta$ do not depend on $I_0$ and $I_1$, correspondingly. The first assumption, $\alpha(I_0)=const$, works when the depth $\delta$ of penetrated magnetic field at sample surface ($\delta = c B_r / 4 \pi J_c$ in CGS units, here $c$ is the velocity of light and $B_r$ is the tangential magnetic field at HTS surface which is twice bigger than PM's field  in "zero approximation"\cite{JAP1}) is much less than system dimensions $L$ (the distance from center of the magnet to sample surface in our geometry), which is the case in the experiment being considered and can always be achieved by choosing appropriate values of $L$ or $\mu$. Moreover, it was shown in Ref.\onlinecite{APL2} that the condition $\delta \ll L$ is too strong and the assumption works well even if $\delta \sim L$. To fulfill the assumption that $\beta(I_1)=const$ is not so easy. In this case, the penetration depth $\delta_{jn}$ of intrajoint field must be less then joint's width that impose strong restrictions on magnetic field value. For the samples which we investigated the joints were $d \sim$ 100 $\mu$m in width that has motivated the readout distances we used to ensure $\delta_{jn} < d$. So, here we measure the superconducting joint's quality for magnetic field 10$^2$--10$^3$ G but the method can be extended to higher fields by replacing levitation force measurements to resonance oscillations technique.\cite{APL1,PhC2}

Another point, which should be mentioned, relates to sample inhomogeneity. Although every method of critical current density determination encounters the same problem a reasonable approximation here is to assume the validity of Eq.\ref{qJ} as integral relation over a region where the main amount of the induced current flows ($\sim 2L$ in our case). This sets a restriction on spatial resolution of the method but also gives a possibility to vary the integrated area using different magnets. For two magnets we used the integrated areas were about 2 mm and 10 mm in diameter respectively and 0.1 mm in depth.

The results of our measurements are summarized in the Table 1. The samples were treated at temperature $T_{tr}$ for 0.17 h in air (1,2) or for 0.5 h in oxygen (3-6). As a solder, the powder of Tm-123 was used for even samples and Tm-123 with addition of 10 wt\% of Y-211 ('green phase') for odd samples. The values $q_{I}$, $q_{II}$ and $q_M$ were determined with first and second magnets and from magnetic flux mapping technique\cite{Prikhna} respectively. The joint's quality value from mapping technique was estimated as $q_M = 2B_{ju}/(B_1+B_2)$, where $B_{ju}$, $B_{1}$, and $B_{2}$ are the local maximum trapped magnetic fields in junction and in two pieces of cut samples. The values of $J_c$ (determined from $F_{un}$, see Ref.\onlinecite{APL2}) and $J_{jn}$ are also represented in the Table. 

Although the mapping technique gives rather rough values it is possible to make some conclusions. First, the $q_{II}$ values are closer to $q_M$ than $q_{I}$  because they are integrated over lager volume. Second, we have compared the average values of quality factors for different series of measurements (here $q \equiv q_I$): over all samples, $\langle q \rangle = 0.57$,  $\langle q_M \rangle = 0.48$; over all in-plane surfaces (parallel to $ab$-plane), $\langle q \rangle_{\parallel} = 0.43$, $\langle q_M \rangle_{\parallel} = 0.50$; and over all out-plane surfaces (perpendicular to $ab$-plane), $\langle q \rangle_{\perp} = 0.70$, $\langle q_M \rangle_{\perp} = 0.47$. Assuming that the quality factors determined from levitation force measurements give the information from a thin undersurface layer (undersurface quality) and the factors determined from mapping represent an integrated values over whole volume (bulk quality), it is possible to say, that the undersurface joint's quality is slightly better than the bulk joint's quality and that the undersurface quality is much better for out-plane surfaces in comparison to in-plane ones. In other words this means that in  the tested samples: (i) the average quality of joints is slightly better at the edges than in the middle; (ii) the critical current density of the material which fills the joints is less anisotropic than the bulk critical current density of the samples itself.

In summary, we have proposed a simple empirical approach to estimate the quality of superconducting joints between blocks of bulk high temperature superconductors. As a measuring value, we introduced a joint's quality factor and found its natural correlation with joint's critical current density. Simultaneous use of the proposed method and flux mapping technique has allowed us to estimate space inhomogeneity of superconducting joints and even obtain information about anisotropy of the joint's critical current density.

\end{document}